# Unpredictability, Uncertainty and Fractal Structures in Physics

Miguel A. F. Sanjuán
Department of Physics
Universidad Rey Juan Carlos
28933 Móstoles, Madrid, Spain
Email: miguel.sanjuan@urjc.es

ABSTRACT

In Physics, we have laws that determine the time evolution of a given physical system, depending on its parameters and its initial conditions. When we have multi-stable systems, many attractors coexist so that their basins of attraction might possess fractal or even Wada boundaries in such a way that the prediction becomes more complicated depending on the initial conditions. Chaotic systems typically present fractal basins in phase space. A small uncertainty in the initial conditions gives rise to a certain unpredictability of the final state behavior. The new notion of basin entropy provides a new quantitative way to measure the unpredictability of the final states in basins of attraction. Simple methods from chaos theory can contribute to a better understanding of fundamental questions in physics as well as other scientific disciplines.

---

The idea of uncertainty has pervaded physics. Among the sources of uncertainty in dynamical systems, we can mention the notion of sensitivity to initial conditions, and the existence of fractal structures in phase space as another one, for a mere simplification.

In this regard, it is interesting to bring up a famous rhyme traditionally associated with Benjamin Franklin (1706-1790), although antecedents of the same idea date back to the 15th century, which is known as *For Want of a Nail* offering an intuitive and poetic image of the idea of sensitive dependence on initial conditions, which is one of the hallmarks of chaos:

> For want of a nail the shoe was lost,
> for want of a shoe the horse was lost,
> for want of a horse the knight was lost,
> for want of a knight the battle was lost,
> for want of a battle the kingdom was lost.
> So a kingdom was lost—all for want of a nail.

Due to the enormous consequences on determinism in physics that quantum mechanics has brought about through Heisenberg uncertainty principle, the idea of indeterminism has been directly related to quantum mechanics. This has led somehow to consider classical mechanics as completely deterministic and predictable, which is not entirely true [1].

It is fascinating to corroborate that the idea of sensitive dependence on initial conditions was considered in detail by the German physicist Max Born (1882-1970), Nobel Prize in Physics in 1954, in an article entitled *Is Classical Mechanics in fact deterministic?* [2]. In it he presented a study of a two-dimensional Lorentz gas initially proposed by the Dutch physicist Hendrik A. Lorentz (1853-1928) in 1905 as a model for the study of electrical conductivity in metals. In this model, a particle moves in a plane that is full of hard spheres and collides with them so that a small change in the initial conditions will significantly alter the trajectory of the particle. This fact led Born to conclude that determinism traditionally related to classical mechanics is not real, since it is not possible to know with infinite precision the initial conditions of a physical experiment.

Furthermore, in the lecture [3] that he gave in 1954 when he received the Nobel Prize we can read the following words:

*"Newtonian mechanics is deterministic in the following sense: If the initial state (positions and velocities of all particles) of a system is accurately given, then the state at any other time (earlier or later) can be calculated from the laws of mechanics. All the other branches of classical physics have been built up according to this model. Mechanical determinism gradually became a kind of article of faith: the world as a machine, an automaton. As far as I can see, this idea has no forerunners in ancient and medieval philosophy. The idea is a product of the immense success of Newtonian mechanics, particularly in astronomy. In the 19th century it became a basic philosophical principle for the whole of exact science. I asked myself whether this was really justified. Can absolute predictions really be made for all time on the basis of the classical equations of motion? It can easily be seen, by simple examples, that this is only the case when the possibility of absolutely exact measurement (of position, velocity, or other quantities) is assumed. Let us think of a particle moving without friction on a straight line between two end-points (walls), at which it experiences completely elastic recoil. It moves with constant speed equal to its initial speed $v_0$ backwards and forwards, and it can be stated exactly where it will be at a given time provided that $v_0$ is accurately known. But if a small inaccuracy $\Delta v_0$ is allowed, then the inaccuracy of prediction of the position at time t is $t\Delta v_0$ which increases with t. If one waits long enough until time $t_c = l/\Delta v_0$ where l is the distance between the elastic walls, the inaccuracy $\Delta x$ will have become equal to the whole space l. Thus, it is impossible to forecast anything about the position at a time which is later than $t_c$. Thus, determinism lapses completely into indeterminism as soon as the slightest inaccuracy in the data on velocity is permitted."*

Likewise, the American physicist Richard Feynman (1918-1988), who won the Nobel Prize for Physics in 1965, makes similar reflections in his well-known book *Lectures in Physics* [4], where he explains that indeterminism is a basic property of many physical systems, and consequently it does not belong exclusively to quantum mechanics.

In the section 38-6 of the first volume of his Lectures in Physics, entitled "Philosophical Implications", a masterful description of indeterminism in classical mechanics is made. The fundamental idea is the uncertainty in accurately setting initial conditions to predict the final state of a physical system. Finally affirming: "Because in classical mechanics there was already indeterminism from a practical point of view".

Precisely another important source of uncertainty in dynamical systems is provided by the fractal structures present in phase space. The natural analogy comes from hydrology, thinking on the basin of a river. A drop of water falling to a river basin goes to the river. We can see geographical maps of river basins dividing a territory of any country in a geographic atlas.

In Nonlinear Dynamics a basin of attraction is defined as the set of initial conditions whose trajectories go to a specific attractor. Furthermore, when we have several attractors in a given region of phase space, we have several basins that are separated by the corresponding boundaries. These boundaries can be classified as smooth basins and fractal basins, depending on the geometrical nature of the boundaries.

In general, we can affirm that when the boundaries are fractal, so that we can also say that the basins are fractal, the fractality implies unpredictability and uncertainty in the future events of trajectories corresponding to the dynamical system associated to theses basins.

An interesting fundamental problem arises when we try to compare a couple of basins, either basins of attraction for dissipative dynamical systems or exit basins for open Hamiltonian systems, since they do not have attractors and as a consequence they cannot have basins of attraction. The fundamental question is to ascertain which basin is more unpredictable. This is the question we may raise by observing the exit basins plotted in Fig.1.

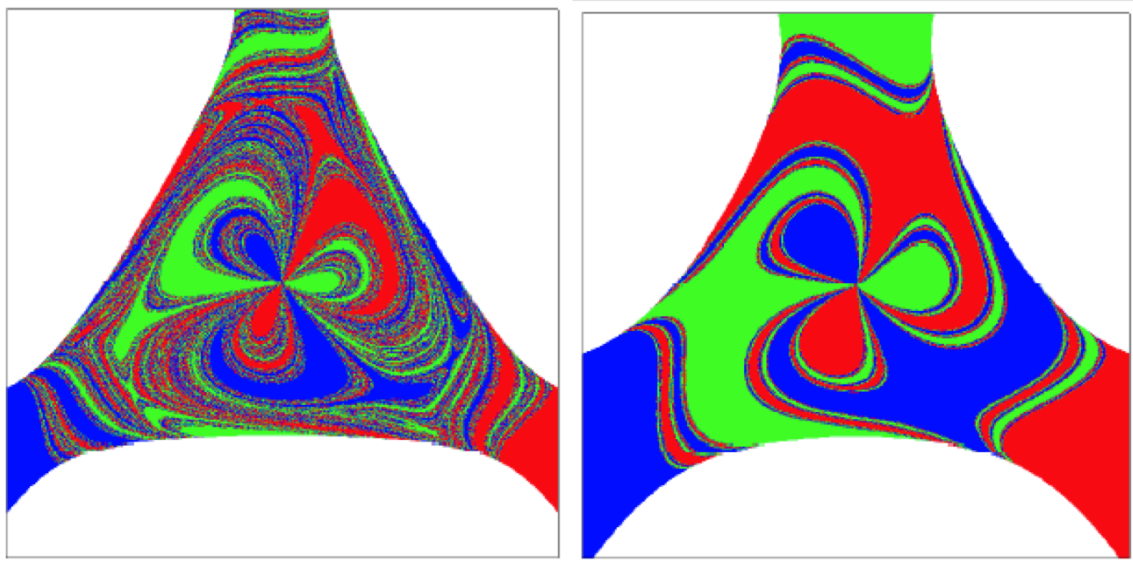

Fig. 1. These figures represent the exit basins corresponding to the Hénon-Heiles Hamiltonian for values of the conserved energy above the critical energy in the physical space, so that the Hamiltonian becomes an open system and three different asymptotic states are possible for orbits whose initial conditions are located in the central region.

Traditionally the unpredictability associated to fractal boundaries has been measured by using the uncertainty dimension. However, there are many examples where we can see that the uncertainty dimension does not help to accurately discriminate among fractal basins with a different degree of unpredictability.

Further for another type of more complicated basins such as riddled basins, where we can say that a basin A is riddled by B, if for every point of A is possible to find arbitrarily close points of B, the uncertainty dimension $\alpha \approx 0$. What basically implies randomness of a deterministic system, and actually two different riddled basins with different structure might not be able to be discerned its degree of unpredictability by using the uncertainty dimension.

Another type of basins are the Wada basins [5], which are fractal basins possessing the Wada property. This property implies that there is a single boundary separating three or more basins, and as a consequence the degree of unpredictability is stronger. For a long period of time there was only one method available to ascertain when a give basin had the Wada property due to Nusse and Yorke [6]. In the past few years, we have developed new methods for testing Wada basins: The Grid Method, the Merging Method and the Saddle-straddle method [7,8,9]. A general overview of how to detect Wada basins is offered in [10].

However, until the appearance of the novel concept of *basin entropy* [11] there was not a quantitative way to identify when a given basin, either Wada or not, was more unpredictable than another one. This is precisely what the basin entropy offers, a quantitative tool to measure the unpredictability of basins. Typically, the algorithm to compute the basin entropy depends on three key ingredients that are related to the size of the boundary, the uncertainty dimension of the basin boundaries and the total number of attractors in the specific region in phase space.

Since the appearance of the concept, it has been applied to numerous problems in physics [12], such as chaotic scattering associated to experiments of cold-atoms [13], chaotic dynamics in relativistic chaotic scattering [14-15], dynamical systems with delay [16], in astrophysics to measure the transition between nonhyperbolic and hyperbolic regimes in open Hamiltonian systems [17], and indirectly through research on Wada structures associated to the dynamics of photons in binary black hole shadows [18] constituting a problem of chaos in general relativity, to cite just a few of them.

The basin entropy quantifies the final state unpredictability of dynamical systems by analyzing the fractal nature of their basins. As such, it constitutes a new tool for the exploration of the uncertainty and unpredictability in nonlinear dynamics. We have applied these methods to different domains in Physics, such as cold atoms, shadows of binary black holes, and classical and relativistic chaotic scattering in astrophysics. We believe that the concept of basin entropy will become an important tool in complex systems studies with applications in multiple scientific fields especially those with multi-stability and other scientific areas as well. Methods derived from nonlinear dynamics have had so far, an enormous influence in many disciplines in science and engineering, though it is important to highlight that tools from the field of chaos theory can be used to understand the rich dynamics of many fundamental problems in physics that are worth to keep exploring through fruitful scientific interactions.